\documentclass[a4paper,11pt]{article}
\pdfoutput=1
\usepackage{jheppub}

\usepackage{amssymb,amsmath}
\usepackage[normalem]{ulem}
\usepackage[utf8x]{inputenc}
\usepackage{slashed}
\usepackage{graphicx}
\usepackage{here}
\usepackage{color}
\usepackage{csquotes} 
\usepackage{comment}
\usepackage{mathrsfs}
\usepackage{float}
\usepackage{ascmac}
\usepackage{multirow}
\usepackage{longtable}

\usepackage[italicdiff]{physics}

\usepackage[hang,small,bf]{caption}
\usepackage[subrefformat=parens]{subcaption}
\usepackage{hyperref}

\makeatletter
\newcommand*\rel@kern[1]{\kern#1\dimexpr\macc@kerna}
\newcommand*\widebar[1]{%
  \begingroup
  \def\mathaccent##1##2{%
    \rel@kern{0.8}%
    \overline{\rel@kern{-0.8}\macc@nucleus\rel@kern{0.2}}%
    \rel@kern{-0.2}%
  }%
  \macc@depth\@ne
  \let\math@bgroup\@empty \let\math@egroup\macc@set@skewchar
  \mathsurround\z@ \frozen@everymath{\mathgroup\macc@group\relax}%
  \macc@set@skewchar\relax
  \let\mathaccentV\macc@nested@a
  \macc@nested@a\relax111{#1}%
  \endgroup
}
\makeatother

\numberwithin{equation}{section}

\preprint{
\begin{minipage}{5cm}
\small
\flushright
KEK-TH-2627
\\
KYUSHU-HET-291
\end{minipage}}

\title{Stabilization of a twisted modulus on a mirror of rigid Calabi-Yau manifold}

\author{Keiya Ishiguro$^{1}$,} 
\author{Takafumi Kai$^{2}$, and} 
\author{Hajime Otsuka$^{2}$} 
\affiliation{
$^1$KEK Theory Center, Institute of Particle and Nuclear Studies, KEK, 1-1 Oho, Tsukuba, Ibaraki 305-0801, Japan}
\affiliation{
$^2$Department of Physics, Kyushu University, 744 Motooka, Nishi-ku, Fukuoka 819-0395, Japan}
\emailAdd{ishigu@post.kek.jp}
\emailAdd{kai.takafumi@phys.kyushu-u.ac.jp}
\emailAdd{otsuka.hajime@phys.kyushu-u.ac.jp}

\abstract{
We study the stabilization of a twisted modulus in Type IIB flux compactifications on a mirror of the rigid Calabi-Yau threefold. By analyzing the effective action of twisted and untwisted moduli, we find that three-form fluxes satisfying the tadpole cancellation conditions lead to supersymmetric AdS vacua. We also investigate swampland conjectures on this non-geometric background.
}

\makeatletter
\gdef\@fpheader{}
\makeatother

\begin{document}

\maketitle

\section{Introduction}

It is important to reveal consistency conditions that the four-dimensional (4D) effective theories admit an ultra-violet completion to a consistent theory of quantum gravity (see for a review, e.g., Ref. \cite{Palti:2019pca}.) 
In the context of swampland program \cite{Vafa:2005ui,ArkaniHamed:2006dz,Ooguri:2006in}, moduli fields appearing in the 4D effective action of string theory are of particular interest in testing the swampland conjectures. 
The vacuum expectation values (VEVs) of moduli fields play important roles in determining the vacuum energy of our universe as well as 4D couplings in the low-energy effective action. 
Type IIB flux compactifications on Calabi-Yau (CY) orientifolds can stabilize all the complex structure moduli and axio-dilaton~\cite{Giddings:2001yu}. 
The remaining K\"ahler moduli will be stabilized at 4D Anti-de Sitter (AdS) minima by utilizing non-perturbative and/or perturbative corrections, as proposed in the Kachru-Kallosh-Linde-Trivedi scenario~\cite{Kachru:2003aw} and Large Volume Scenario \cite{Balasubramanian:2005zx}. 
However, the validity of de Sitter (dS) vacua uplifted by certain non-perturbative effects and/or an existence of anti D3-brane is still an important open question~\cite{Gautason:2018gln,Bena:2018fqc,Blumenhagen:2019qcg,Gao:2020xqh}.

In this paper, we focus on a different class of CY manifold, the so-called ``non-geometric'' CY manifold with vanishing K\"ahler moduli, i.e., $h^{1,1}=0$. 
This background geometry will allow a rigorous check of swampland conjectures. 
In particular, we deal with a mirror of a rigid CY manifold. 
Such a rigid CY manifold with $h^{2,1}=0$ was used in Type IIA flux compactifications, e.g., DeWolfe-Giryavets-Kachru-Taylor (DGKT) model \cite{DeWolfe:2005uu}. 
Since the mirror of the rigid CY manifold does not have the K\"ahler deformations, one can only focus on the dynamics of complex structure moduli in Type IIB side. 
The stabilization of complex structure on such a background geometry has been studied in Refs. \cite{Becker:2006ks,Becker:2007dn,Ishiguro:2021csu,Bardzell:2022jfh,Becker:2022hse} in several moduli space of the complex structure moduli, but the vacuum structure of both the untwisted and twisted moduli is not fully explored.\footnote{Recently, there has been an attempt for the stabilization of all the complex structure moduli around the Fermat point \cite{Becker:2024ijy}.} 
The purpose of this paper is to provide a method to derive the effective action of twisted moduli by utilizing a mirror symmetry technique in the context of Type IIB flux compactifications. 
With this prescription, we study the flux compactification of complex structure moduli and verify swampland conjectures such as the AdS/moduli scale separation conjecture and species scale distance conjecture. 
Our numerical analysis shows that three-form fluxes satisfying the tadpole cancellation condition lead to supersymmetric AdS vacua which are consistent with swampland conjectures.

This paper is organized as follows. 
In Sec. \ref{sec:setup}, we briefly review Type IIB flux compactifications on the mirror of the rigid CY manifold, following Refs. \cite{Candelas:1990pi,Becker:2006ks,Becker:2007dn,Ishiguro:2021csu}. 
By utilizing the special geometry of the underlying manifold, we write down the flux-induced superpotential, i.e., Gukov-Vafa-Witten (GVW) type superpotential \cite{Gukov:1999ya}, as a function of untwisted and twisted moduli fields. 
In Sec. \ref{sec:swampland}, we verify swampland conjectures for the VEVs of moduli fields obtained in Sec. \ref{sec:setup}. 
Finally, Sec. \ref{sec:con} is devoted to the conclusions.

\section{Setup}
\label{sec:setup}

In this section, we derive the effective action of both untwisted and twisted moduli in Type IIB flux compactifications. 
In Sec. \ref{sec:geometry}, we briefly review the geometric structure of background geometry. 
In Sec. \ref{sec:period}, we show the period vector including the contribution of a twisted modulus whose asymptotic expansion is shown in Sec. \ref{sec:asymptotic}. The effective action in the context of Type IIB flux compactifications is discussed in Sec. \ref{sec:eft}.

\subsection{Geometry}
\label{sec:geometry}

We begin with the $1^9$ Gepner model \cite{Gepner:1987qi,Lutken:1988zj,Greene:1988ut} which is known as to the ${\cal Z}$ manifold with $b_{11}({\cal Z})=36$ and $b_{21}({\cal Z})=36$~\cite{Candelas:1985en,Strominger:1985it}. 
The Landau-Ginzburg potential is described by
\begin{align}
    W = \sum_{k=1}^9 y_k^3\,,
\end{align}
where $y_k$ is regarded as a homogeneous coordinate of $\mathbb{P}_8$. 

It was known that a mirror of the rigid CY manifold ${\cal Z}$ is described by a quotient of $\mathbb{P}_8[3]$ with a certain group $G$, i.e., $\tilde{\cal Z}= \mathbb{P}_8[3]/G$. 
Although this background geometry is a seven-dimensional manifold, 
one can consider the same cohomology structure of usual CY threefolds \cite{Candelas_1993}. 
Indeed, the Hodge number of $H^7(\tilde{\cal Z}) = \oplus_{p+q=7} H_{\Bar{\partial}}^{p,q}(\tilde{\cal Z})$ is given by $(h^{0,7},h^{1,6},h^{2,5},h^{3,4},h^{4,3},h^{5,2},h^{6,1},h^{7,0})=(0,0,1,\beta,\beta,1,0,0)$ with $\beta$ being 84 for $\mathbb{P}_8[3]$. 
Hence, the middle cohomology ($H^{2,5}, H^{3,4}, H^{4,3}, H^{5,2}$) has the same structure for the Hodge decomposition of $H^3$ for CY threefolds. 
Furthermore, existence of a unique $(5,2)$-form indicates that the complex structure moduli space is described by special geometry, as an analogue of the unique holomorphic three-form of CY threefolds.

In this paper, we focus on a mirror of the rigid CY manifold $\tilde{\cal Z}=\mathbb{P}_8[3]/\mathbb{Z}_3$ whose defining equation is given by
\begin{align}
    W = \sum_{i,j=1}^3 x_{ij}^3 - 3 \sum_{k=1}^3 \phi_k e_k - 3 \sum_{m,n,p=1}^3 s_{mnp} f_{mnp}\,,
\end{align}
where $e_i \simeq x_{i1}x_{i2}x_{i3}$ and $f_{mnp}\simeq x_{1m}x_{2n}x_{3p}$ with $x_{ij}=y_{3i+j-3}$. Here, $\phi_k$ and $s_{mnp}$ correspond to untwisted and twisted moduli fields, respectively. 
Note that when we turn off the twisted moduli, i.e., $s_{mnp}=0$ $\forall m,n,p$, the background geometry is described by three factorizable tori, each which is defined on $\mathbb{P}_2[3]$. 
In this paper, we turn on one of the twisted moduli fields as an illustrative purpose. It corresponds to a generalization of the analysis of Ref. \cite{Ishiguro:2021csu} which deals with only untwisted moduli. 
On top of that, we introduce the orientifold action with the tadpole charge 12 \cite{Becker:2006ks}. 
Thanks to the symplectic structure of the background geometry, one can introduce background fluxes in the context of Type IIB flux compactifications. 

Before going into the detail of flux compactifications, we describe special geometry on a generalized CY manifold ${\cal M}$. Let us consider a symplectic basis $(A^a, B_b)$ of $H_7({\cal M},\mathbb{Z})$ with $a, b = 0, 1,..., b_{4,3}+1$.
They satisfy the following relations:
\begin{align}
    A^a \cap B_b = \delta^a_b, \quad B_b \cap A^a = - \delta^a_b, \quad A^a \cap A^b = 0 \quad B_a \cap B_b = 0.
\end{align}
The dual cohomology basis $(A^a, B_b)$ of $H_7({\cal M},\mathbb{Z})$ is defined such that it satisfies 
\begin{align}
    \int_{A^a} \alpha_b = \int_{\cal M} {\alpha_b \wedge \beta^a} = \delta^a_b, \quad \int_{B_a} \beta^b = \int_{\cal M} {\beta^b \wedge \alpha_a} = - \delta^b_a.
\end{align}
Then, we introduce the so-called period vector as follows:
\begin{align}
    \Pi \equiv \begin{pmatrix}
        F_a \\
        u^a
    \end{pmatrix}
    \equiv \begin{pmatrix}
        \int_{B_a} \Omega \\
        \int_{A^a} \Omega
    \end{pmatrix}
    .
\end{align}
The projective coordinates $u^a$ are defined on moduli space by using an integral of the $(5,2)$-form $\Omega$ over the $A^a$-cycle, and 
$F_a$ is a function of $u^a$ which is also defined by the corresponding integral. 
Hence, the $(5,2)$-form $\Omega$ is expanded by
\begin{align}
    \Omega = u^a \alpha_a - F_a \beta^a.
    \label{eq:omega}
\end{align}
The number of projective coordinates defined this way is $h^{4,3} + 1$.
However, the coordinates $u^a$ are only defined up to a complex rescaling.
Taking into account this factor, we consider the quotient:
\begin{align}
    \tau_{\alpha} = \frac{u^{\alpha}}{u^0} \qquad \alpha =1,..., h^{4,3},
\end{align}
where the index $\alpha$ excludes 0 from the index $a$.
By setting $u^0 = 1$, $\tau_{\alpha}$ becomes a set of dynamical fields, namely the complex structure moduli, and this way gives the right number of coordinates to describe the moduli.

\subsection{The periods for twisted moduli}
\label{sec:period}
In this research, we examine a specific period vector calculated in Ref. \cite{Candelas_1993}.
Suppose we let only one of the twisted moduli be non-zero, the period vector is given by
\begin{align}
    \Pi \equiv \begin{pmatrix}
        F_0 \\
        F_1 \\
        F_2 \\
        F_3 \\
        F_4 \\
        u^0 \\
        u^1 \\
        u^2 \\
        u^3 \\
        u^4
    \end{pmatrix}
    =f
    \begin{pmatrix}
    -\varpi_{QQQ} \\
    \varpi_{RQQ} \\
    \varpi_{QRQ} \\
    \varpi_{QQR} \\
    c(\omega^2 \varpi + \omega \hat{\varpi}) \\
    \varpi_{RRR} \\
    \varpi_{QRR} \\
    \varpi_{RQR} \\
    \varpi_{RRQ} \\
    c(\varpi + \hat{\varpi})
    \end{pmatrix},
\end{align}
where $\omega = e^{\frac{2\pi i}{3}}, ~ c = \frac{(2 \pi)^6 i}{3^{5/2}}$ and the gauge factor $f = \prod_{i=1}^3 J^{- \frac{1}{3}} (\phi_i) \left( \frac{d\phi_i}{dJ} \right)^{1/2}$ with $J(\phi_i)=\frac{\phi_i^3}{4^3} \frac{(\phi_i^3 + 8)^3}{(\phi_i^3 - 1)^3}$. 
Here and in what follows, we define the untwisted moduli $\phi_i$ and twisted modulus $\chi$, respectively. 
Let us introduce 
\begin{align}
    \varpi_{IJK} = \sum_{r = 0}^\infty \frac{(3\chi)^{3r}}{(3r)!} I(\phi_1, r)J(\phi_2, r)K(\phi_3, r)
    ,
\end{align}
for $I, J, K = \{Q, R\}$, and $Q$ and $R$ are functions depending on $\phi_i$ via $Z_1, Z_2$ as
\begin{align}
    \begin{aligned}
         Q(\phi_i, r) &= \frac{(2\pi)^2}{3}(- Z_1(\phi_i, r) + Z_2 (\phi_i, r)), \\
         R(\phi_i, r) &= \frac{(2\pi)^2}{3} (- i \sqrt{3}) (\omega^2 Z_1(\phi, r) + \omega Z_2 (\phi, r)),
    \end{aligned}
    \label{eq:Q_and_R}
\end{align}
with
\begin{align}
    \begin{alignedat}{2}
        Z_1(\phi_i, r) &= &\frac{\Gamma(\frac{1}{3}) \Gamma(r + \frac{1}{3})}{\Gamma(\frac{2}{3})} \ _{2}F_{1}\left(\frac{1}{3},r+\frac{1}{3};\frac{2}{3}; \phi_i^3\right), \\
        Z_2(\phi_i, r) &= \phi_i &\frac{\Gamma(\frac{2}{3}) \Gamma(r + \frac{2}{3})}{ \Gamma(\frac{4}{3})} \ _2F_1 \left(\frac{2}{3}, r + \frac{2}{3}; \frac{4}{3}; \phi_i^3 \right).
    \end{alignedat}
\end{align}
The 5- and 10-th components $\varpi, \hat{\varpi}$ in the period vector are given by
\begin{align}
    \begin{aligned}
        \varpi=\sum_{r=0}^{\infty} \frac{(3 \chi)^{3 r+1}}{(3 r+1) !} Z_{3}\left(\phi_{1}, r\right) Z_{3}\left(\phi_{2}, r\right) Z_{3}\left(\phi_{3}, r\right), \\
        \widehat{\varpi}=\sum_{r=0}^{\infty} \frac{(3 \chi)^{3 r+2}}{(3 r+2) !} Z_{4}\left(\phi_{1}, r\right) Z_{4}\left(\phi_{2}, r\right) Z_{4}\left(\phi_{3}, r\right),
    \end{aligned}
\end{align}
with 
\begin{align}
    \begin{aligned}
        Z_{3}(\phi_i, r) &= Z_{1}\left(\phi_i, r+\frac{1}{3}\right), \\
        Z_{4}(\phi_i, r) &= Z_{2}\left(\phi_i, r+\frac{2}{3}\right).
    \end{aligned}
\end{align}

Now we consider the limit $\chi \rightarrow 0$ that a set of eight periods of the ten basis elements reduce to the periods of three torus.
In this limit, $\varpi_{IJK}$ is given by
\begin{align}
    \varpi_{IJK} &\simeq I(\phi_1, 0)J(\phi_2, 0)K(\phi_3, 0) \qquad \left( I, J, K = \{Q, R\} \right).
\end{align}
Also, $\varpi, \hat{\varpi}$ are given by
\begin{align}
    \begin{aligned}
        \varpi &\simeq 3 \chi Z_{3}\left(\phi_{1}, 0\right) Z_{3}\left(\phi_{2}, 0\right) Z_{3}\left(\phi_{3}, 0\right) \\
        &= 3 \chi \prod_{i=1}^3 Z_{1}\left(\phi_{i}, \frac{1}{3}\right), \\
        \widehat{\varpi} &\simeq \frac{(3 \chi)^2}{2} Z_{4}\left(\phi_{1}, 0\right) Z_{4}\left(\phi_{2}, 0\right) Z_{4}\left(\phi_{3}, 0\right) \\
        &= \frac{(3 \chi)^2}{2} \prod_{i=1}^3 Z_{2}\left(\phi_{i}, \frac{2}{3}\right).
    \end{aligned}
    \label{eq:varpi_and_varpihat}
\end{align}
The functions of $Q$ and $R$ have important relations for the untwisted complex structure moduli as follows:
\begin{align}
    \tau_{\alpha} = \frac{Q (\phi_{\alpha})}{R (\phi_{\alpha})} \qquad (\alpha = 1, 2, 3),
\end{align}
Hence, we obtain a specific period vector that has three moduli $\tau_i ~ (i = 1, 2, 3)$ and twisted modulus $\chi$. We arrive at the period vector:
\begin{align}
    \Pi =
    \begin{pmatrix}
    - \tau_1 \tau_2 \tau_3 \\
    \tau_2 \tau_3 \\
    \tau_1 \tau_3 \\
    \tau_1 \tau_2 \\
    c f (\omega^2 \varpi + \omega \hat{\varpi}) \\
    1 \\
    \tau_1 \\
    \tau_2 \\
    \tau_3 \\
    c f (\varpi + \hat{\varpi})
    \end{pmatrix},
\end{align}
where $f = \frac{1}{R(\phi_1)R(\phi_2)R(\phi_3)}$.

\subsection{Asymptotic approximation of hypergeometric function}
\label{sec:asymptotic}

In this section, we examine the asymptotic approximation of hypergeometric function which appears in the period vector. 
The hypergeometric function $_2F_1$ has a following expansion:
\begin{align}
    &_2F_1(a, a+m; c; z) 
   = \frac{\Gamma(c)}{\Gamma(c-a) \Gamma(a + m)} (-z)^{-a-m} \sum_{n=0}^{\infty} \frac{(a)_{n + m} (a- c + 1)_{n + m}}{n! (n+m)!} \nonumber \\ & \times z^{-n} [\log(-z) + \psi(1+n) - \psi(a + n + m) - \psi(c - a - n - m) + \psi(1 + n + m)] \nonumber \\ & + \frac{\Gamma(c)}{\Gamma(a+m)} (-z)^{-a} \sum_{n=0}^{m-1} \frac{(a)_n \Gamma(m-n)}{n! \Gamma(c-a-n)}z^{-n},
\label{eq:2F1}
\end{align}
with $|{\rm arg}(- z)|\leq \pi, |z| > 1, c-a \notin \mathbb{Z}, m \in \mathbb{N} \cup \{0\}$.
Here, $\Gamma(x)$ is a gamma function, $\psi(x)$ is a polygamma (digamma) function and $(a)_n = \Gamma(a + n)/\Gamma(a)$.

To apply this expression of the hypergeometric function into some elements of the period vector, it is notable that $\phi$ and $\tau$ are asymptotically related as follows.\footnote{For the moment, we omit the index $i$ of $\phi_i$ and $\tau_i$.} From the relation \cite{Candelas_1993}:
\begin{align}
    J (\tau) = \frac{\phi^3}{4^3} \frac{(\phi^3 + 8)^3}{(\phi^3 - 1)^3}. \qquad \left( J \sim \frac{1}{12^3} e^{- 2\pi i \tau} \right),
\end{align}
$\phi$ and $\tau$ are related as $\phi \sim \frac{1}{3} e^{- \frac{2\pi i \tau}{3}}$. 
To satisfy this approximation $\phi \sim \frac{1}{3} e^{- \frac{2\pi i \tau}{3}}$, it becomes apparent that at least the following lower bound for $\text{Im} \tau$ is necessary:
\begin{align}
    \text{Im} \tau > \frac{\text{log} 729}{2 \pi}.
\end{align}
Furthermore, $J (\tau)$ can be explicitly described by using $q$-expansion as follows:
\begin{align}
    12^3 J (\tau) = \frac{1}{q} + 744 + \mathcal{O} (q) \qquad (q \equiv e^{2\pi i \tau}).
\end{align}
Then, similarly to the above, it becomes necessary to impose further lower bounds on $\text{Im} \tau$ to fulfill $J \sim \frac{1}{12^3} e^{- 2\pi i \tau}$:
\begin{align}
    \text{Im} \tau > \frac{\text{log} 744}{2 \pi}.
\end{align}
Therefore, our analysis to explore flux vacua is valid for the range $\text{Im} \tau \geq 1.1$.

Then, in the large complex structure regime $\phi \sim \frac{1}{3} e^{- \frac{2\pi i \tau}{3}} = \frac{1}{3} e^{\frac{2 \pi {\rm Im}\tau}{3}} e^{- \frac{2\pi i {\rm Re}\tau}{3}} \rightarrow \infty$ with ${\rm Arg} \phi = - \frac{2\pi {\Re}\tau}{3} (0 \leq {\rm Re} \tau <1)$, $Z_1, Z_2$ appearing in the period vector are given by 
\begin{align}
    \begin{aligned}
        Z_1 (\phi, 0) &\sim (- \phi^3)^{- \frac{1}{3}}\left[\log(-\phi^3) - 2\psi\left(\frac{1}{3}\right) + 2\gamma\right],\\
        Z_2 (\phi, 0) &\sim (- \phi^3)^{- \frac{2}{3}} \left[\log(-\phi^3) - 2\psi\left(\frac{2}{3}\right) + 2\gamma\right], \\
    \end{aligned}
    \label{eq:Z1_and_Z2}
\end{align}
where $z = - \phi^3$ and $\gamma$ is a Euler's constant. 
It was known in Ref. \cite{Candelas_1993} that the background geometry enjoys the $\Pi_i SL(2,\mathbb{Z})_i$ modular symmetry associated with three untwisted complex structure moduli $\tau_i$. 
For a fundamental domain of ${\rm Re}\,\tau_i$, for instance, we restrict ourselves to the region of $0 \leq {\rm Re}\,\tau_i < 1$ because of $|{\rm arg}(-\phi_i^3 ) |\leq \pi$, but the other region of ${\rm Re}\,\tau_i$ is also possible to analyze.\footnote{By considering $T_i$ transformations of $SL(2, \mathbb{Z})_i$ and the redefinition of $\tau_i$, it is possible to consider a transition to different branches of the hypergeometric function. Therefore, in the range of $\text{Re}\,\tau_i < 0, 1 \leq \text{Re}\,\tau_i$, the asymptotic expressions of the hypergeometric function in Eq. (\ref{eq:2F1}) can also be utilized.} 
Then, the asymptotic expression of function $R$ is derived as follows: 
\begin{align}
    \begin{aligned}
         R (\phi_i, 0) = &-i \frac{(2 \pi)^2}{\sqrt{3}} \left( \omega^2 Z_1 (\phi_i, 0) + \omega Z_2 (\phi_i, 0) \right) \\ 
         \sim &-i \frac{(2 \pi)^2}{\sqrt{3}} \frac{1}{\phi_i} \left[ 2 \psi \left( \frac{1}{3} \right) - 2  \psi \left( \frac{2}{3} \right) \right].
    \end{aligned} 
    \label{eq:R_function}
\end{align}

Similarly, for the 5-th and 10-th directions of the period vector, we can obtain the explicit expression of $\varpi, \hat{\varpi}$:
\begin{align}
    \begin{alignedat}{2}
        \varpi &\sim 3 \chi \prod_{i=1}^3 \frac{\Gamma \left( \frac{1}{3} \right)}{(1 - \phi_i^3)^{1/3}} & &\sim 3 \chi \prod_{i=1}^3 \frac{\Gamma \left( \frac{1}{3} \right)}{(-1)^{1/3} \phi_i}, \\
        \widehat{\varpi} &\sim \frac{(3 \chi)^2}{2} \prod_{i=1}^3 \phi_i \frac{\Gamma \left( \frac{2}{3} \right)}{(1 - \phi_i^3)^{2/3}} & &\sim \frac{(3 \chi)^2}{2} \prod_{i=1}^3 \frac{\Gamma \left( \frac{2}{3} \right)}{(-1)^{2/3} \phi_i},
    \end{alignedat}
\end{align}
where $_2F_1 (a, b; b; z) = (1 - z)^{-a}$ and $|\phi_i| \rightarrow \infty$. Therefore, in the limit $\text{Im} \tau \rightarrow \infty$, the 5- and 10-th components $\Pi_5, \Pi_{10}$ in the period vector are given by
\begin{align}
    \begin{aligned}
        \Pi_5 &= cf ( \omega^2 \varpi + \omega \hat{\varpi} ) \\
        &\sim \frac{1}{24} \left( \psi \left( \frac{1}{3} \right) - \psi \left( \frac{2}{3} \right) \right)^{-3} \left[ -3 \chi \omega^2 \Gamma^3 \left( \frac{1}{3} \right) + \frac{(3 \chi)^2}{2} \omega \Gamma^3 \left( \frac{2}{3} \right) \right], 
        \\
        \Pi_{10} &= cf ( \varpi + \hat{\varpi} ) \\
        &\sim \frac{1}{24} \left( \psi \left( \frac{1}{3} \right) - \psi \left( \frac{2}{3} \right) \right)^{-3} \left[ -3 \chi \Gamma^3 \left( \frac{1}{3} \right) + \frac{(3 \chi)^2}{2} \Gamma^3 \left( \frac{2}{3} \right) \right].
    \end{aligned}
    \label{eq:pi5_and_pi10}
\end{align}

\subsection{Effective action of untwisted and twisted moduli in Type IIB flux compactification}
\label{sec:eft}

Here, we discuss Type IIB flux compactifications on the mirror of the rigid CY manifold, following  Refs. \cite{Becker:2007dn, Ishiguro:2021csu}. 
The three-form fluxes $F_3$ and $H_3$ are quantized on each cycle which 
are defined as
\begin{align}
    \begin{alignedat}{2}
        N_{F_a} &\equiv \int_{A^a} F_3, & \quad N_{F_{a+5}} &\equiv \int_{B^a} F_3, \\
        N_{H_a} &\equiv \int_{A^a} H_3, & \quad N_{H_{a+5}} &\equiv \int_{B^a} H_3,\quad (a = 0,..., 4) 
        \label{eq:fluxes}
    \end{alignedat}
\end{align}
so that $\{ N_{F_0},..., N_{F_9}, N_{H_0},..., N_{H_9} \}$ becomes a set of integers.
When these fluxes are turned on three cycles, GVW superpotential in the 4D effective action is provided by \cite{Gukov:1999ya} 
\begin{align}
    W = \int_{\tilde{\cal{Z}}} G_3 \wedge \Omega,
\end{align}
with $G_3 \equiv F_3 - S H_3$. Here in what follows, the 4D reduced Plank mass is set to be 1, and $S$ is defined by $S \equiv C_0 + i e^{-\phi}$ with $C_0$ and $\phi$ being the Ramond-Ramond 0-form and 4D dilaton, respectively.
Then, we can find an explicit form of the superpotential by taking into account Eqs. \eqref{eq:omega}, \eqref{eq:pi5_and_pi10} and \eqref{eq:fluxes} as follows:
\begin{align}
    \begin{aligned}
        W &= (N_{F_0} - S N_{H_0})\tau_1 \tau_2 \tau_3 \\
        &- \left[ (N_{F_1} - S N_{H_1}) \tau_2 \tau_3 + (N_{F_2} - S N_{H_2})\tau_3 \tau_1 + (N_{F_3} - S N_{H_3})\tau_1 \tau_2 \right] \\
        &- (N_{F_4} - S N_{H_4}) \Pi_5 \\
        &+ (N_{F_5} - S N_{H_5}) \\
        &+ \left[ (N_{F_6} - S N_{H_6}) \tau_1 + (N_{F_7} - S N_{H_7})\tau_2 +(N_{F_8} - S N_{H_8})\tau_3 \right] \\
        &+ (N_{F_9} - S N_{H_9}) \Pi_{10}.
    \end{aligned}
    \label{eq:explicit_superpotential}
\end{align}

The K\"{a}hler potential in the case of geometric CY threefolds is given by 
\begin{align}
    K = K_{\text{ad}} + K_{\text{cs}} + K_{\text{vol}} = -\text{log}(-i(S - \bar{S})) - \text{log} \left( -i \int \Omega \wedge \bar{\Omega} \right) - 2 \text{log} \mathcal{V},
\end{align}
where $\mathcal{V}$ is the volume of CY threefolds. 
However, in the non-geometric case, the K\"{a}hler moduli has no deformation under the orbifolding as discussed in Ref. \cite{Becker:2007dn}. 
The contribution of fixed K\"{a}hler moduli modifies the axio-dilaton K\"{a}hler potential:
\begin{align}
    K_{\text{ad}} + K_{\text{vol}} = - 4\text{log}(-i(S - \bar{S})),
    \label{eq:axio-dilaton_Kahler}
\end{align}
As regards the K\"{a}hler potential of the complex structure moduli, we employ $K_{\text{cs}}$ as the potential including up to the second order of twisted modulus:
\begin{align}
    \begin{aligned}
        K_{\text{cs}} &= - \text{log} \left(i \prod_{i=1}^3 (\tau_i - \bar{\tau}_i) - i \left( \Pi_5 \bar{\Pi}_{10} - \Pi_{10} \bar{\Pi}_5 \right) \right) \\
        &\sim - \text{log} \left(i \prod_{i=1}^3 (\tau_i - \bar{\tau}_i)  \right) + \frac{(2 \pi)^9 \pi^3 \Gamma^6 \left( \frac{1}{3} \right)}{3^{5/2}} \frac{|\chi|^2}{|R_1 R_2 R_3|^2} \prod_{i=1}^3 \frac{|1 - \phi_i^3|^{-\frac{2}{3}}}{\text{Im} \tau_i} \\
        &\sim - \text{log} \left(i \prod_{i=1}^3 (\tau_i - \bar{\tau}_i)  \right) + A |\chi|^2 \prod_{i=1}^3 \frac{1}{\text{Im} \tau_i}.
    \end{aligned}
    \label{eq:CS_Kahler}
\end{align}
Then, by taking into account Eq. (\ref{eq:R_function}), we can estimate the value of $A$ as follows:
\begin{align}
    A  = \left( \frac{(2 \pi)^2}{\sqrt{3}} \left[ 2 \psi \left( \frac{1}{3} \right) - 2 \psi \left( \frac{2}{3} \right) \right] \right)^{-6} \frac{(2 \pi)^9 \pi^3 \Gamma^6 \left( \frac{1}{3} \right)}{3^{5/2}},
\end{align}
As a result, the 4D scalar potential $V$ is defined in terms of $K$ and $W$:
\begin{align}
    V = e^K ( K^{I \bar{J}} D_I W D_{\bar{J}} \bar{W} - 3 |W|^2 ),
\end{align}
where $D_I W \equiv W \partial_I K + \partial_I W$, $K_{I \bar{J}} \equiv \partial_I \partial_{\bar{J}} K$ denotes the K\"{a}hler metric, and the index $I$ runs the complex structure moduli and the axio-dilaton.

Finally, we discuss the D3-brane charge that appears in the 4D effective action. The $N_{\text{flux}}$ constituted by the 3-form fluxes is defined as follows:
\begin{align}
    \begin{aligned}
        N_{\text{flux}} &= \int_{\tilde{\cal Z}} H_3 \wedge F_3 \\
        &= N_{F_5} N_{H_0} + N_{F_6} N_{H_1} + N_{F_7} N_{H_2} + N_{F_8} N_{H_3} + N_{F_9} N_{H_4} \\
        &\quad - N_{F_0} N_{H_5} - N_{F_1} N_{H_6} - N_{F_2} N_{H_7} - N_{F_3} N_{H_8} - N_{F_4} N_{H_9}.
    \end{aligned}
    \label{eq:N_flux}
\end{align}
Moreover, $N_{\text{flux}}$ is canceled by the number of D3-branes ($N_{\text{D3}}$) and O3-planes ($N_{\text{O3}}$) according to the tadpole cancellation condition:
\begin{align}
    N_{\text{flux}} + N_{\text{D3}} - \frac{1}{2} N_{\text{O3}} = 0.
\end{align}
Given that the specific values of O3-planes are determined by orientifold actions in Ref. \cite{Becker:2006ks}, the maximum value of $N_{\text{flux}}$ is restricted to 12.
Hence, the following upper bound exists for $N_{\text{flux}}$:
\begin{align}
    N_{\text{flux}} \leq 12.
\end{align}

\section{Numerical analysis and swampland conjectures}
\label{sec:swampland}

In Sec. \ref{sec:AdS}, we first numerically analyze the effective action. 
For the obtained SUSY AdS vacua, we next examine the AdS/moduli scale separation conjecture in Sec. \ref{sec:conjecture1}. 
Finally, the species scale and distance conjecture on this non-geometric background are discussed in Sec. \ref{sec:conjecture2}.

\subsection{SUSY AdS/Minkowski vacua with twisted moduli}
\label{sec:AdS}
We numerically analyze the scalar potential by using the superpotential \eqref{eq:explicit_superpotential} and the Kähler potential \eqref{eq:axio-dilaton_Kahler} and \eqref{eq:CS_Kahler}. 
To search for the flux vacua of scalar potential numerically, we utilize the “FindRoot” function in Mathematica. 
In this research, we examine the flux vacua within the specified range for the three-form flux quanta:
\begin{align}
    -10 \leq \{ N_{F_0},..., N_{F_9}, N_{H_0},..., N_{H_9} \} \leq 10.
    \label{eq:range_of_fluxes}
\end{align}
Considering all possible combinations of three-form fluxes within the range of Eq. (\ref{eq:range_of_fluxes}), there are approximately $10^{15}$ sets of fluxes in the isotropic case $\tau:=\tau_1=\tau_2=\tau_3$ and $10^{26}$ sets in the anisotropic case.
Note that in the isotropic case, we assume the flux relations $N_{F_1} = N_{F_2} = N_{F_3}$, $N_{F_6} = N_{F_7} = N_{F_8}$, $N_{H_1} = N_{H_2} = N_{H_3}$ and $N_{H_6} = N_{H_7} = N_{H_8}$. 
To simplify our numerical analysis, we randomly generate sets of fluxes within the range of Eq. (\ref{eq:range_of_fluxes}).

Here, we will discuss the necessary constraints involved in the random exploration of stable vacua.
As discussed in Sec. \ref{sec:asymptotic}, the obtained flux vacua adhere to the constraint $\text{Im} \tau \geq 1.1$.
Considering the limit as $\chi \rightarrow 0$ in Eq. (\ref{eq:varpi_and_varpihat}), we discuss the restriction $|\chi| < 1$ on the VEVs of the twisted modulus. 
Moreover, since we consider the effective supergravity action in perturbative string theory, it is necessary to focus on the weak coupling region $\langle\text{Im} S\rangle = g_{\text{s}}^{-1} > 1$.
Although it seems to be no lower bound for $N_{\rm flux}$, we constrain the range of fluxes using $|N_{\rm flux}| \leq 12$.

\begin{table}[H]
    \centering
    \begin{tabular}{|c|c|c|}\hline
         &  Minkowski ($W=0$) & AdS ($W\neq0$)  \\ \hline
       Isotropic case with $|N_{\rm flux}| \leq 12$  & 0 $\quad (10^5)$ & $42923 \quad (10^6)$ \\ \hline
       Anisotropic case with $|N_{\rm flux}| \leq 12$  & 0 $\quad (10^5)$ & $648 \quad (2 \times 10^{6})$ \\ \hline
    \end{tabular}
    \caption{The numbers of stable vacua, where the number in the parentheses denotes the generated set of three-form fluxes.}
    \label{tab:number_vacua}
\end{table}

In Table \ref{tab:number_vacua}, we summarize the number of stable vacua for a given set of fluxes.
It turns out that in the random search for flux vacua using Eqs. (\ref{eq:explicit_superpotential}) (\ref{eq:axio-dilaton_Kahler}) (\ref{eq:CS_Kahler}), no SUSY Minkowski solutions are found in the isotropic and anisotropic cases, and only SUSY AdS solutions are allowed. \footnote{Even if we generate $10^5$ flux sets that are restricted to $G_3 \in H^{4,3}$ in the isotropic and anisotropic cases, we cannot find SUSY Minkowski solutions.} 
For illustrative purposes, we show the benchmark points for supersymmetric AdS vacua in Table \ref{tab:vacua_value}, where we list the values of the three-form fluxes as well as the D3-brane charge induced by fluxes $N_{\rm flux}$, and the VEVs of moduli fields and the scalar potential. 
Note that these vacua are classically stable, and we also present the mass squared of the lightest modulus $m^2_{\text{light}}$ in the untwisted and twisted sectors. 
Moreover, these $m^2_{\text{light}}$ and the scalar potential in Table \ref{tab:number_vacua} respectively satisfy that $m_{\rm light}^2 < m_{\rm KK}^2$ and $\langle V \rangle < m_{\rm KK}^4$, where the KK mass is given by Eq. \eqref{eq:KKmass1}.
Then, the number of SUSY AdS solutions in the isotropic case satisfying the above restrictions is only 44, and we cannot find any solutions in the anisotropic case under the restrictions.

\begin{table}[H]
    \centering
    \begin{tabular}{cccc}\hline
       Properties  &  Vacuum 1 & Vacuum 2 & Vacuum3 \\ \hline
       $N_{F_0}$  & -5 & 1 & 2 \\ 
       $N_{F_1}$  & 1 & 0 & 0 \\ 
       $N_{F_4}$  & 7 & 2 & 6 \\ 
       $N_{F_5}$  & 1 & 6 & -10 \\
       $N_{F_6}$  & -3 & 1 & 0 \\ 
       $N_{F_9}$  & 10 & 7 & -4 \\ 
       $N_{H_0}$  & 0 & 0 & 0 \\ 
       $N_{H_1}$  & 1 & 0 & 0 \\ 
       $N_{H_4}$  & 4 & -1 & -9 \\ 
       $N_{H_5}$  & -4 & 0 & 4 \\ 
       $N_{H_6}$  & -1 & -2 & 1 \\ 
       $N_{H_9}$  & 2 & -7 & 3 \\ 
       $\langle \text{Re} S \rangle$  & 0.524 & -0.0519 & -1.87 \\
       $\langle \text{Im} S \rangle$  & 2.61 & 1.74 & 1.49 \\
       $\langle \text{Re} \tau \rangle$  & -0.0507 & -0.335 & 0.0743 \\
       $\langle \text{Im} \tau \rangle$  & 1.31 & 1.30 & 1.25 \\
       $\langle \text{Re} \chi \rangle$  & $2.80 \times 10^{-8}$ & $1.08 \times 10^{-8}$ & $8.21 \times 10^{-9}$ \\
       $\langle \text{Im} \chi \rangle$  & $2.14 \times 10^{-8}$ & $-1.43 \times 10^{-8}$ & $-2.73 \times 10^{-8}$ \\
       $\langle V\rangle$  & -0.00543 & -0.0576 & -0.0430 \\
       $N_{\text{flux}}$  & 0 & 7 & 10 \\
       $m^2_{\text{light}}$  & 0.0253 & 0.108 & 0.0260\\
       \hline
    \end{tabular}
    \caption{Some explicit values of SUSY AdS vacua in the isotropic case.}
    \label{tab:vacua_value}
\end{table}

\subsection{AdS/moduli scale separation conjecture}
\label{sec:conjecture1}
For the obtained supersymmetric AdS vacua, we will examine the swampland conjecture. 
In particular, we focus on the overall untwisted moduli, i.e., $\tau=\tau_1=\tau_2=\tau_3$.
The AdS/moduli scale separation conjecture posits that in the AdS minimum, the size of the AdS space cannot be separated from the lightest mass of the moduli~\cite{Gautason:2018gln}.
In this context, the following relation exists between the size of the AdS and the lightest modulus:
\begin{align}
    m_{\text{light}} R_{\text{AdS}} \leq c,
\end{align}
where $c$ is $\mathcal{O}(1)$ constant.
The mass of the lightest modulus satisfies $m_{\text{light}} \sim R_{AdS_5}$, and $AdS_5 \times S^5$ solution of type IIB superstring upholds this conjecture.
Note that the 5-form fluxes are related to the sizes of $R_{\text{AdS}}$ and $R_{S^5}$.

\begin{figure}[H]
\centering
\includegraphics[width = 0.8 \linewidth]{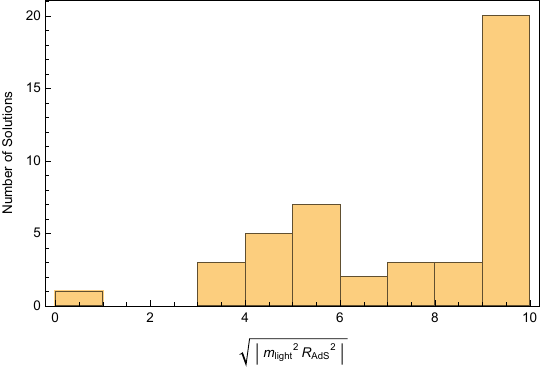}
\caption{The value of $\sqrt{|m^2_{\text{light}} R^2_{\text{AdS}}|}$ at SUSY AdS vacua in the isotropic moduli space.}
\label{fig:AdS/moduli_scale_separation}
\end{figure}

In the following, we will discuss the AdS/moduli scale separation conjecture numerically.
Here, we employ the following expression as the specific $R_{\text{AdS}}$ in Ref. \cite{Ishiguro:2021csu}:
\begin{align}
    R_{\text{AdS}} = \sqrt{\frac{(d - 1) (d - 2)}{|\Lambda_{\text{AdS}}|}}.
    \label{eq:rAdS}
\end{align}
Using Eq. \eqref{eq:rAdS} and the numerical values obtained from the random search of flux vacua, we summarize the distribution of $\sqrt{|m_{\text{light}}^2 R_{\text{AdS}}^2|}$ in Fig \ref{fig:AdS/moduli_scale_separation}.

It turns out that this distribution peaks at $\sqrt{|m_{\text{light}}^2 R_{\text{AdS}}^2|} \sim 9$.
On the other hand, the results in Ref. \cite{Ishiguro:2021csu} have peaks at the different value of $\sqrt{|m_{\text{light}}^2 R_{\text{AdS}}^2|}$.
According to this result, we can see that the value of $\sqrt{|m_{\text{light}}^2 R_{\text{AdS}}^2|}$ is affected to be larger when moduli stabilization is performed including contributions from twisted sectors.
In contrast to the SUSY AdS vacua, an analysis of the relation between $\sqrt{|m_{\text{light}}^2 R_{\text{AdS}}^2|}$ and $N_{\rm flux}$ for the sets of VEVs in the isotropic case of Table \ref{tab:number_vacua} shows that the dependence of $N_{\rm flux}$ does not appear in $\sqrt{|m_{\text{light}}^2 R_{\text{AdS}}^2|}$.

\subsection{Species scale and distance conjecture}
\label{sec:conjecture2}

Lastly, we examine the species scale $\Lambda_s$ on the non-geometric CY manifold. 
In the presence of a large number of light particles, the cutoff scale in the 4D EFT is smaller than the Planck scale. 
This scale is called the species scale \cite{Dvali:2007hz,Dvali:2009ks,Dvali:2010vm,Dvali:2012uq}. 
Through the Kaluza-Klein (KK) compactifications, the light degrees of freedom correspond to string and KK modes, and the species scale is given by the string scale for the light string case and the higher-dimensional Planck mass for the light KK case \cite{Agmon:2022thq}. 
Indeed, in an infinite-distance limit in the moduli space, it was proposed in Ref. \cite{Lee:2019xtm} that there are two types of light tower of states in string theory: (i) a tower of light string states corresponding to an emergent string limit at which a charged fundamental string becomes tensionless, and (ii) a tower of light KK states corresponding to a decompactification limit.

In the light string scale, the scale of higher-derivative corrections in the 4D effective action is controlled by the string scale, and the species scale is regarded as a cutoff scale. 
Then, the species scale is determined by the string scale:
\begin{align}
    \Lambda_s^{\rm string} = M_s \sim  g_s M_{\rm Pl}.
\end{align}
When we approach the boundary of the moduli space of dilaton, a light tower of states appears in the 4D effective action, as known in the distance conjecture \cite{Ooguri:2006in}. 
 In a large distance in the moduli space, the typical mass scale in units of 4D reduced Planck mass behaves as 
\begin{align}
    m \sim e^{-\lambda \phi},
\end{align}
where $\phi$ denotes the canonically normalized modulus, and $\lambda\sim {\cal O}(1)$ in units of $M_{\rm Pl}=1$.

In weakly-coupled perturbative string theory on CY threefolds, 
the K\"ahler potential of the 4D axio-dilaton is $K= -\ln (i (\Bar{S}-S))$, and 
the canonical normalization of the 4D dilaton $\phi_s = \ln({\rm Im}\,S)/\sqrt{2}$ 
leads to 
\begin{align}
    \Lambda_s^{\rm string} \simeq e^{-\frac{\phi_s}{\sqrt{2}}} M_{\rm Pl},
\end{align}
with $\lambda = 1/\sqrt{2}$. 
Note that in the case of a fundamental string, the tension scales as $T_{\rm string}\sim e^{-\frac{2}{\sqrt{d-2}}\phi_s}$ in the $d$-dimensional theory. (For more details, see, e.g., Ref. \cite{Etheredge:2022opl}.)

In contrast to the geometric CY compactifications, the K\"ahler potential of the axion dilaton is modified as in \eqref{eq:axio-dilaton_Kahler} on the non-geometric background. 
Thus, the canonical normalization of the 4D dilaton $\phi_s =\sqrt{2} \ln({\rm Im}\,S)$ changes the dilaton dependence of the species scale to 
\begin{align}
    \Lambda_s^{\rm string} \simeq e^{-\frac{\phi_s}{2}} M_{\rm Pl},
\end{align}
with $\lambda = \frac{1}{2}$. 
It seems to violate the bound on $\lambda$, i.e., $\lambda \geq \frac{1}{\sqrt{d-2}}$ proposed in Ref. \cite{Etheredge:2022opl}, but this bound is applicable to the lightest tower of states in an infinite-distance limit of the moduli space. Indeed, the tower of KK mass is lighter than these string states, as will be shown later. 
On top of that, $\lambda=\frac{1}{2}$ satisfies the bound $\lambda \geq \lambda_{\rm min} = 1/\sqrt{(d-1)(d-2)}$ proposed in $d$ dimensions from several theoretical viewpoints \cite{Grimm:2018ohb,Andriot:2020lea,Gendler:2020dfp,Lanza:2020qmt,Bedroya:2020rmd,Lanza:2021udy,Calderon-Infante:2023ler}.

The KK mass can be extracted from the T-dual Type IIA side. 
By using the expression of KK mass $m_{\rm KK}$ in Eq. (60) of Ref. \cite{Blumenhagen:2019vgj}, one can evaluate the lightest KK mass on the mirror of CY manifold:
\begin{align}
    m^2_{\rm KK} \simeq \frac{1}{({\rm Im}S)^2 {\rm Im \tau}_{\rm max}}, 
\label{eq:KKmass1}
\end{align}
with ${\rm Im \tau}_{\rm max} \equiv {\rm Max}\{{\rm Im \tau}_1, {\rm Im \tau}_2, {\rm Im \tau}_3\}$ for the anisotropic case and ${\rm Im \tau}_{\rm max} \equiv {\rm Im \tau}$ for the isotropic case. 
Here, we assume that the vacuum expectation value of twisted modulus is smaller than that of untwisted modulus, and we show the KK mass associated with the largest untwisted cycle. 
Hence, the lightest KK mass is smaller than the string scale since the KK mass has an additional suppression factor with respect to the complex structure modulus. 
After canonically normalizing fields, the lightest KK mass is estimated as
\begin{align}
    m^2_{\rm KK} \simeq e^{-\frac{\phi_s}{2} - \sqrt{2}\phi_\tau},
\label{eq:KKmass2}
\end{align}
where $\phi_\tau =\ln({\rm Im}\,\tau)/\sqrt{2}$ denote the canonically normalized complex structure modulus. Here, we consider the anisotropic case, and $\phi_\tau =\sqrt{\frac{3}{2}}\ln({\rm Im}\,\tau)$ in the isotropic case. 
When there are several infinite towers of light states in a certain infinite-distance limit, $\lambda$ is taken to be the largest one, as discussed in the convex hull swampland distance conjecture \cite{Calderon-Infante:2020dhm}. 
For the lightest KK mass \eqref{eq:KKmass2}, the maximum value of $\lambda$, i.e., $\lambda= 1/\sqrt{2}$, saturates the bound $\lambda \geq 1/\sqrt{d-2}$. 

\section{Conclusions}
\label{sec:con}

In this paper, we have studied the stabilization of both the untwisted and twisted moduli on the mirror of the rigid CY manifold, as an extension of Ref. \cite{Ishiguro:2021csu}. 
In this class of background geometry, three-form fluxes can stabilize all the geometric moduli fields due to the lack of K\"ahler moduli, and it will be a moderate background geometry to verify swampland conjectures. 

We present the method to calculate the Type IIB effective action of both untwisted and twisted moduli by utilizing the period vector developed in Ref. \cite{Candelas:1990pi}. 
With this prescription, one can write down the flux-induced potential as a function of twisted moduli, as shown in Sec. \ref{sec:setup}. 
When we turn on a single twisted modulus, we find that three-form fluxes within the tadpole cancellation condition lead to the stabilization of all of the moduli at supersymmetric AdS vacua. 
Furthermore, it is possible to realize a small contribution to the tadpole such as $N_{\rm flux}=0$. 
We restricted ourselves to a single twisted modulus, but it is interesting to explore the vacuum structure of such a mirror of the rigid CY manifold, which is left for future work. 

The K\"ahler potential of the axio-dilaton is different from the usual geometric CY compactifications due to the fixed K\"ahler moduli. 
Hence, it is interesting to check the proposed swampland conjectures. 
In particular, we focused on the AdS/moduli scale separation conjecture and species scale distance conjecture. 
Our numerical analysis exhibits that the parameter in the AdS/moduli scale separation conjecture is indeed ${\cal O}(1)$ in a similar to the previous analysis \cite{Ishiguro:2021csu}, and ${\cal O}(1)$ parameter is peaked around a specific value. 
Furthermore, we examined the species scale for two types of light tower of states in string theory, i.e., a tower of light string states and light KK states. 
In the case of light string states, the species scale, i.e., the string scale has a novel dependence on the dilaton which leads to the small $\lambda$ in the species scale $\Lambda_s \sim e^{-\lambda \phi}$ in contrast to geometric CY compactifications, but it satisfies the bound on the minimum value for $\lambda_{\rm min}= 1/\sqrt{(d-1)(d-2)}$. 
In the case of light KK states, they satisfy the bound on $\lambda$, $\lambda \geq 1/\sqrt{d-2}$, since the lightest tower of states corresponds to KK states in the large complex structure regime of untwisted moduli.

{\bf Note added}

After finishing this work, we learned of another work~\cite{Becker:2024ijy} where the stabilization of twisted moduli around the Fermat point was studied.

\acknowledgments

This work was supported in part by Kyushu University’s Innovator Fellowship Program (T.K.) and JSPS KAKENHI Grant Numbers JP23H04512 (H.O).

\bibliography{references}{}
\bibliographystyle{JHEP}

\end{document}